



\font\titlefont = cmr10 scaled\magstep 4
\font\sectionfont = cmr10
\font\littlefont = cmr5
\font\eightrm = cmr8

\def\ss{\scriptstyle}
\def\sss{\scriptscriptstyle}

\newcount\tcflag
\tcflag = 0  

\ifnum\tcflag = 0 \magnification = 1200 \fi  

\global\baselineskip = 1.2\baselineskip
\global\parskip = 4pt plus 0.3pt
\global\abovedisplayskip = 18pt plus3pt minus9pt
\global\belowdisplayskip = 18pt plus3pt minus9pt
\global\abovedisplayshortskip = 6pt plus3pt
\global\belowdisplayshortskip = 6pt plus3pt


\def\endignore{}
\def\ignore #1\endignore{}

\newcount\dflag
\dflag = 0


\def\monthname{\ifcase\month
\or January \or February \or March \or April \or May \or June%
\or July \or August \or September \or October \or November %
\or December
\fi}

\newcount\dummy
\newcount\minute  
\newcount\hour
\newcount\localtime
\newcount\localday
\localtime = \time
\localday = \day

\def\advanceclock#1#2{ 
\dummy = #1
\multiply\dummy by 60
\advance\dummy by #2
\advance\localtime by \dummy
\ifnum\localtime > 1440 
\advance\localtime by -1440
\advance\localday by 1
\fi}

\def\settime{{\dummy = \localtime%
\divide\dummy by 60%
\hour = \dummy
\minute = \localtime%
\multiply\dummy by 60%
\advance\minute by -\dummy
\ifnum\minute < 10
\xdef\spacer{0} 
\else \xdef\spacer{}
\fi %
\ifnum\hour < 12
\xdef\ampm{a.m.} 
\else
\xdef\ampm{p.m.} 
\advance\hour by -12 %
\fi %
\ifnum\hour = 0 \hour = 12 \fi
\xdef\timestring{\number\hour : \spacer \number\minute%
\thinspace \ampm}}}



\def\endtitle{}
\def\title#1\endtitle{\vskip.5in\titlefont
\global\baselineskip = 2\baselineskip
#1\vskip.4in
\baselineskip = 0.5\baselineskip\rm}

\def\endauthors{}
\def\authors#1\endauthors{#1}

\def\endabstract{}
\def\abstract#1\endabstract{\vskip .3in%
\centerline{\sectionfont\bf Abstract}%
\vskip .1in
\noindent#1}

\def\nopageonenumber{\footline={\ifnum\pageno<2\hfil\else
\hss\tenrm\folio\hss\fi}}  

\newcount\nsection
\newcount\nsubsection

\def\section#1{\global\advance\nsection by 1
\nsubsection=0
\bigskip\noindent\centerline{\sectionfont \bf \number\nsection.\ #1}
\bigskip\rm\nobreak}

\def\subsection#1{\global\advance\nsubsection by 1
\bigskip\noindent\sectionfont \sl \number\nsection.\number\nsubsection)\
#1\bigskip\rm\nobreak}

\def\topic#1{{\medskip\noindent $\bullet$ \it #1:}}
\def\endtopic{\medskip}

\def\appendix#1#2{\bigskip\noindent%
\centerline{\sectionfont \bf Appendix #1.\ #2}
\bigskip\rm\nobreak}


\newcount\nref
\global\nref = 1

\def\therefs{}


\def\ref#1#2{\xdef #1{[\number\nref]}
\ifnum\nref = 1\global\xdef\therefs{\item{[\number\nref]} #2\ }
\else
\global\xdef\oldrefs{\therefs}
\global\xdef\therefs{\oldrefs\vskip.1in\item{[\number\nref]} #2\ }%
\fi%
\global\advance\nref by 1
}

\def\listrefs{\vfill\eject\section{References}\therefs}


\newcount\nfoot
\global\nfoot = 1

\def\foot#1#2{\xdef #1{(\number\nfoot)}
\footnote{${}^{\number\nfoot}$}{\eightrm #2}
\global\advance\nfoot by 1
}


\newcount\nfig
\global\nfig = 1
\def\thefigs{} 

\def\figure#1#2{\xdef #1{(\number\nfig)}
\ifnum\nfig = 1\global\xdef\thefigs{\item{(\number\nfig)} #2\ }
\else
\global\xdef\oldfigs{\thefigs}
\global\xdef\thefigs{\oldfigs\vskip.1in\item{(\number\nfig)} #2\ }%
\fi%
\global\advance\nfig by 1 } 

\def\figurecaptions{\vfill\eject\section{Figure Captions}\thefigs}

\def\fig#1{\xdef #1{(\number\nfig)}
\global\advance\nfig by 1 } 


\newcount\cflag
\newcount\nequation
\global\nequation = 1
\def\eqlabel{(1)}

\def\nexteqno{\ifnum\cflag = 0
\global\advance\nequation by 1
\fi
\global\cflag = 0
\xdef\eqlabel{(\number\nequation)}}

\def\lasteqno{\global\advance\nequation by -1
\xdef\eqlabel{(\number\nequation)}}

\def\label#1{\xdef #1{(\number\nequation)}
\ifnum\dflag = 1
{\escapechar = -1
\xdef\draftname{\littlefont\string#1}}
\fi}

\def\clabel#1#2{\xdef\eqlabel{(\number\nequation #2)}
\global\cflag = 1
\xdef #1{\eqlabel}
\ifnum\dflag = 1
{\escapechar = -1
\xdef\draftname{\string#1}}
\fi}

\def\cclabel#1#2{\xdef\eqlabel{#2)}
\global\cflag = 1
\xdef #1{\eqlabel}
\ifnum\dflag = 1
{\escapechar = -1
\xdef\draftname{\string#1}}
\fi}


\def\eeq{}

\def\eqnn #1\eeq{$$ #1 $$}

\def\eq #1\eeq{
\ifnum\dflag = 0
{\xdef\draftname{\ }}
\fi 
$$ #1
\eqno{\eqlabel \rlap{\ \draftname}} $$
\nexteqno}







\def\eqa #1\eeq{
\ifnum\dflag = 0
{\xdef\draftname{\ }}
\fi 
$$ \eqalignno{ #1 } $$
\global\cflag = 0}


\def\ie{{\it i.e.\/}}

\def\etc{{\it etc.\/}}
\def\etal{{\it et.al.\/}}
\def\apriori{{\it a priori\/}}


\def\mpla#1#2#3{{\it Mod.\ Phys.\ Lett.} {\bf A#1}, (19#2) #3}

\def\npb#1#2#3{{\it Nucl.\ Phys.} {\bf B#1} (19#2) #3}
\def\plb#1#2#3{{\it Phys.\ Lett.} {\bf #1B} (19#2) #3}

\def\prd#1#2#3{{\it Phys.\ Rev.} {\bf D#1} (19#2) #3}

\def\prl#1#2#3{{\it Phys.\ Rev.\ Lett.} {\bf #1} (19#2) #3}


\global\nulldelimiterspace = 0pt



\def\frac#1#2{{{#1} \over {#2}}\,}  



\def\Dsl{\hbox{/\kern-.6700em\it D}} 
\def\dsl{\hbox{/\kern-.5300em$\partial$}}
\def\pxpsl{\hbox{/\kern-.5600em$p$}}
\def\ssl{\hbox{/\kern-.5300em$s$}}
\def\epssl{\hbox{/\kern-.5100em$\epsilon$}}
\def\delsl{\hbox{/\kern-.6300em$\nabla$}}
\def\lxpsl{\hbox{/\kern-.4300em$l$}}
\def\elxpsl{\hbox{/\kern-.4500em$\ell$}}
\def\kxpsl{\hbox{/\kern-.5100em$k$}}
\def\qxpsl{\hbox{/\kern-.5000em$q$}}
\def\sla#1{\raise.15ex\hbox{$/$}\kern-.57em #1}



\def\roughly#1{\mathrel{\raise.3ex\hbox{$#1$\kern-.75em
  \lower1ex\hbox{$\sim$}}}}





\def\Scl{{\cal L}}


\def\ssl{{\sss L}}

\def\ssr{{\sss R}}

\def\ssw{{\sss W}}

\def\ssz{{\sss Z}}


\def\Re{{\rm Re\;}}






\def\GeV{{\rm \ GeV}}

\overfullrule=0pt


\def\nc{{\rm nc}}

\def\rht{{\sss R}}
\def\lft{{\sss L}}
\def\sw{s_w}
\def\cw{c_w}
\def\Mw{M_{\sss W}}
\def\Mz{M_{\sss Z}}

\def\mH{m_{\sss H}}
\def\mt{m_t}

\def\as{\alpha_s(M_{{\sss Z}})}
\def\GF{G_{\sss F}}

\def\leff{\Scl_{\rm eff}}

\def\gl{g_\lft}
\def\gr{g_\rht}

\def\zff{Zf\overline{f}}
\def\zbb{Zb\overline{b}}

\def\ALR{A_{\sss LR}}
\def\AFB#1{A^0_{\sss FB}(#1)}
\def\AbFB{A_{{\sss FB}}(b)}
\def\AcFB{A_{{\sss FB}}(c)}
\def\Gtot{\Gamma_{{\sss Z}}}
\def\UD{{\sss UD}}
\def\dUD{\delta_\UD}

\voffset0.5in

\line{hepph/9505339 \hfill McGill-95/18, NEIP-95-005, UTTG-09-95}


\title
\centerline{New Physics and Recent High}
\centerline{Precision Electroweak Measurements}
\endtitle

\authors
\centerline{P.~Bamert${}^a$, C.P. Burgess${}^b$ and Ivan Maksymyk${}^c$}
\vskip .1in
\centerline{\it ${}^a$ Institut de Physique, Universit\'e de Neuch\^atel}
\centerline{\it CH-2000 Neuch\^atel, Switzerland.}
\vskip .05in
\centerline{\it ${}^b$ Physics Department, McGill University}
\centerline{\it 3600 University St., Montr\'eal, Qu\'ebec, CANADA, H3A
2T8.}
\vskip .05in
\centerline{\it ${}^c$ Theory Group, Department of Physics}
\centerline{\it University of Texas, Austin, TX 78712.}
\endauthors

\abstract
We analyze LEP and SLC data from the 1995 Winter Conferences for signals
of new physics.  We compare the data with the Standard Model (SM) as well
as a number of test hypotheses concerning the nature of new physics:
($i$) nonstandard $\zbb$ couplings, ($ii$) nonstandard $\zff$ couplings for the
entire
third generation, ($iii$) nonstandard oblique corrections, ($iv$) nonstandard
lepton couplings, ($v$) general nonstandard $W$ and $Z$ couplings to all
fermions,
as well as combinations of the above. In most of our analyses, we leave the SM
variables $\alpha_s$ and $m_t$ as free parameters to see how the various types
of new physics can affect their inferred values. We find that the best fit
($\chi^2$/d.o.f.
= 8.4/10) is obtained for the nonstandard $\zbb$ couplings, which also give a
`low'
value (0.112) for $\alpha_s$.  The SM also gives a good description of the $Z$
data,
having $\chi^2$/d.o.f. = 12.4/12. If $\alpha_s$ is held fixed to the low-energy
value
0.112, then we find that a combination of the nonstandard $\zbb$ couplings is
fit to lie more than four standard deviations away from zero.
\endabstract


\vfill\eject
\section{Introduction}

Recently announced results from LEP (Moriond 1995)
boast an overall energy error of 1.5 MeV
in the measurement of the $Z$-mass (\ie\ one part in $10^5$), and
an error of roughly one part in $10^3$ for $Z$-decay rates and
branching ratios. In fact, the LEP experiments are even
sensitive to such small gravitational effects as the tidal forces due to the
sun and the moon, as well as due to changing water levels in Lake Geneva.
Moreover,  the experimental error for the SLC measurement
of the asymmetry $A_{{\sss LR}}$ is a remarkable 0.4\%.
Such high precision permits very
rigorous tests of the standard model (SM) of the
strong and electroweak interactions. With this precise data,
one hopes to find discrepancies between experiment
and SM predictions, and to thereby gain an indication of
the kind of the new physics which will ultimately prevail in its stead.

\ref\oblique{B. Lynn, M. Peskin, R. Stuart, in {\it Physics at LEP}, CERN
Report 86-02.}
\ref\holdom{B. Holdom and J. Terning, \plb{247}{90}{88}.}
\ref\STU{M.E. Peskin and T. Takeuchi, \prl{65}{90}{964}; \prd{46}{92}{381};
W.J. Marciano and J.L. Rosner, \prl{65}{90}{2963};
D.C. Kennedy and P. Langacker, \prl{65}{90}{2967}.}
\ref\ABC{G. Altarelli and R. Barbieri, \plb{253}{91}{161};
G. Altarelli, R. Barbieri and S. Jadach, \npb{369}{92}{3}, (erratum)
{\it ibid.} {\bf B376} (1992) 444;
G. Altarelli, R. Barbieri and F. Caravaglios, \npb{405}{93}{3}.}
\ref\STUVWX{I. Maksymyk, C.P. Burgess and D. London, \prd{50}{94}{529};
C.P Burgess, S. Godfrey, H. K\"onig, D. London and I. Maksymyk,
\plb{326}{94}{276}.}
\ref\TR{T. Takeuchi, A.K. Grant and J.L. Rosner, preprint
Fermilab-Conf-94-279-T
(hep-ph/9409211).}
\ref\bigfit{C.P. Burgess, S. Godfrey, H. K\"onig, D. London and I. Maksymyk,
\prd{49}{94}{6115}.}

This letter reports on our most recent analysis of Z-pole data for indications
of
new physics. This analysis differs from those that have been done previously in
two important ways: ($i$) it includes the most recent results reported by
the experimental groups in the 1995 winter conferences, and ($ii$) it tests a
large number of hypotheses of new physics, and is not limited to a
consideration of
the specific forms which have often been used in the past (such as `oblique'
\oblique,
\holdom, \STU, \ABC, \STUVWX\ and/or new $\zbb$ couplings \ABC, \TR). We are
able to perform a broader survey of the theoretical possibilities by taking
advantage of the effective-lagrangian approach recently proposed in
ref.~\bigfit.
This approach entails only a bare minimum of theoretical prejudice,
our goal being to give the data as free a hand as possible to indicate in which
direction new physics lies.

We organize the presentation of our results as follows. We first briefly
review the general
effective lagrangian used to parametrize new physics effects
at the $Z$ pole. This is followed by a summary of
the most recent experimental data. We then compare the data with
several test hypotheses concerning the nature of the new physics,
considering a wide variety of choices of combinations
of effective couplings. This comparison permits a quantitative statistical
evaluation of the relative compatibility of each hypothesis
with the data. Our conclusions are
finally summarized in the last section.

\section{Parameterizing the New Physics}

Based on the effective-lagrangian approach outlined in \bigfit, the present
analyses incorporates all of the potential effective operators which can arise
up to and including dimension four. This involves three different kinds of
nonstandard interactions. The first kind consists of nonstandard contributions
to electroweak boson propagation, as can be parameterized using the usual
oblique parameters  $S$ and $T$. Next, there are non-standard neutral-current
$\zff$ interactions, which we normalize according to
\eq\label\leffdirect
\leff^{\nc} =
-\; {e\over \sw \cw} \overline{\psi}_f \gamma^\mu \left[ \left( \gl^f + \delta
\gl^f
\right) \gamma\lft +  \left(\gr^f  + \delta \gr^f \right) \gamma\rht \right]
\psi_f,
\eeq
where the SM couplings are given in terms of the fermion's weak isospin and
electric charge by $\gl^f =  I_3^f - Q^f \sw^2 $ and $\gr^f = - Q^f \sw^2$.
$\sw
= \sin\theta_\ssw$ where $\theta_\ssw$ is the usual weak mixing angle.
Finally come nonstandard fermion-$W$ couplings, whose strength we parameterize
by $\delta h_\ssl^{ff'}$ and $\delta h_\ssr^{ff'}$, where the normalization is
such
that the SM contribution for leptons is $h_\ssl^{ff'} = \delta_{ff'}$.
It is an easy matter to derive predictions for observables in terms of these
parameters \bigfit, and these
are most usefully cast as a SM prediction supplemented by a new-physics
correction
which is linearized in the new couplings $S$, $T$, $\gl^f$, \etc.

All three classes of nonstandard interactions can contribute
to precision measurements taken purely at the $Z$ pole, although the
nonstandard
fermion-$W$ interactions only appear through the specific combination
$\Delta  = \sum_{f=e,\mu} \left[ \sqrt{\sum_i \left| \delta_{if} + \delta
h_\lft^{\nu_i f} \right|^2} \; -1 \right] \approx \Re(\delta h_L^{\nu_e e  }) +
\Re(\delta h_L^{\nu_\mu \mu}) $. Only this combination appears because
nonstandard
fermion-$W$ couplings play a role solely through their contribution to muon
decay, from which the measured value of the Fermi coupling, $\GF$, is inferred.
This value is relevant since $\GF$ (together, of course, with
$\alpha$ and $\Mz$), are used as inputs to define the SM predictions for the
$Z$-pole observables.

At the outset, it therefore appears that new physics can affect the $Z$-pole
observables through 19 new-physics parameters: $\{ \delta g_{{\sss L,R}}^f,
S,T, \Delta \}$, with $f=e, \mu, \tau, u, d, s, c ,b$.
It is not possible, however, to simultaneously constrain all these
parameters from $Z$-pole data since they do not all appear independently
in the measured quantities. For example, $Z$-pole observables only depend
on the quantities $T$ and $\Delta$ through the combination $\alpha T -
\Delta$ \bigfit. Because of this we ignore $\Delta$ in our fits, although all
of
our results for $T$ are properly interpreted as constraints for $\alpha T -
\Delta$. Similarly, since the hadron-related observables (which
we take to be $\AbFB$, $\AcFB$, $\Gtot$,  $R_l \equiv \Gamma_{\rm had}/
\Gamma_{l}$ (for $l=e,\mu,\tau$), $\sigma_h \equiv 12\pi \Gamma_{\rm had}
\Gamma_{e}/\Mz^2 \Gtot^2$) only depend on the light-quark couplings through the
hadronic width, $\Gamma_{\rm had}$, the only measurable combination of these
couplings is
\eq\label\UDdef
\dUD \equiv \sum_{q=u,d,s} \Bigl( \gl^q \delta \gl^q + \gr^q \delta \gr^q
\Bigr).
\eeq
It must also be noted that neither of the oblique parameters, $S$ and $T$, can
be separated from the nonstandard $\zff$ couplings, if such new couplings are
permitted for all species of fermions.  Additional information becomes
available
once other experimental quantities, such as $\Mw$ or low-energy scattering
cross sections, are also considered. None of the fits reported here avail
themselves
of this additional information, however.

In analyzing the data we test a variety of assumptions concerning the nature of
new physics, by imposing \apriori\ relations amongst the nonstandard couplings.
We imagine a series of cases ranging from the SM only to a fit in which
as many constrainable parameters as possible are allowed to float.
Intermediate
cases include permitting only nonstandard $\zbb$ couplings,
$\delta g^b_{\ssl,\ssr}$; permitting nonstandard $\zff$ couplings for the
entire
third generation, $\delta g^b_{\ssl,\ssr}, \delta g^\tau_{\ssl,\ssr}$;
permitting
only oblique corrections, $S$ and $T$; permitting light-lepton couplings,
$g^e_{\ssl, \ssr}, g^\mu_{\ssl,\ssr}$; as well as various combinations of these
alternatives.

\section{Data Analysis and Discussion}

\ref\LEP{The LEP Electroweak Working Group, {\it A Combination of Preliminary
LEP Electroweak Results for the 1995 Winter Conferences}, preprint
LEPEWWG/95-01,
ALEPH 95-038, DELPHI 95-37, L3 Note 1736, OPAL Note TN284. }
\ref\slc{K. Abe et al., \prl{73}{94}{25}, (hep-ex/9404001).}

The 1995 winter conferences saw the release \LEP\ of new numbers from the
LEP experiments for observables on the $Z$ pole. The numbers for those
observables which we use in our analysis, together with the most recent figure
from SLC \slc\ for $\ALR$, are listed in Table I. Although we do not list them
explicitly here, we take the experimental correlations for the LEP observables
from ref.~\LEP.

We have fit these observables using a variety of assumptions concerning the
nature of new physics. We give a representative set of the results which come
from
these fits in Table II through VIII.  For Tables III through VIII, which
describe the
results of fits which include the influence of new physics, the column labelled
`Pull'
indicates the deviation from zero of the best-fit value for each new-physics
parameter,
in units of the standard deviation for that parameter. This quantifies the
extent with
which the data prefer these nonstandard couplings to be nonzero.

A few other points of explanation concerning these tables are in order.

\topic{(1) The SM Fit}
Table II lists the values for the strong coupling constant, $\as$, and the
top-quark
mass, $\mt$, that are obtained by fitting
the data of Table I to the SM predictions. We choose the fiducial value
$\mH = 300$ GeV for the Higgs mass in all of our fits, and ignore the
comparatively
weak dependence of the observables on $\mH$. This fit is meant to verify that
our
fits reproduce the usual results for $\as$ and $\mt$ when they are restricted
to the
SM case, as well as to establish a baseline against which to compare the
quality
of the fits to extensions of the SM.

 \ref\top{F. Abe \etal, \prl{74}{95}{2626}, (hep-ex/9503002);
S. Abachi \etal, \prl{74}{95}{2632}, (hep-ex/9503003).}

\topic{(2) Sensitivity to $\mt$}
We  find that, irrespective of the types of new
physics which are assumed, fits in which $\mt$ floats
generally give values of $\mt$ that are clustered around
the value obtained from our SM fit (174$\pm$ 8 GeV for fits
based on LEP data only,
180$\pm$ 7 GeV for fits based on LEP data and SLC).
Thus, leaving $\mt$ free to float in fits including
new physics does not
appreciably improve agreement with the
data.  Moreover, we see that the new physics we
consider does not ruin the agreement between the value for $\mt$ that
is inferred from $Z$ physics, and that which has been found from the
kinematics of $t$ production at Fermilab \top.

Since the dominant $\mt$ dependence (\ie\ those one-loop corrections
proportional to $\GF \mt^2$)
can be considered to be contributions to $T$ and $\delta \gl^b$, one
combination
of $\mt$, $T$ and $\delta \gl^b$ becomes
poorly constrained when all three of these parameters are free
to float in a fit. In practice the same is approximately true if it is only $T$
and
$\mt$ that float since most observables depend only weakly on $\delta \gl^b$.
We therefore fix $\mt$ in those fits which involve the oblique parameters,
choosing $\mt = 179$ GeV, although we find our results do not depend strongly
on the value chosen for $\mt$ (within a few decades of 175 GeV).

\topic{(3) Inclusion of $\ALR$}
We have performed these fits with and without the SLC result for $\ALR$.
As can be seen from the tables, the inclusion of $\ALR$ always
deteriorates the quality of the fit. This is because $\ALR$ and the LEP
observables
independently measure, and give differing values for, the same combination
($A_e$)
of electron couplings. Although new physics of the type we consider can change
the value of $A_e$ from the SM prediction, it appears in $\ALR$ and the LEP
measurements of $A_e$ in the same way, and so cannot reconcile the two
sets of experimental results. We interpret this to indicate the likely
existence of
an as yet unidentified source of systematic error in one or the other of these
experiments. Until the source of this error is found, we quote our results both
with
and without the inclusion of $\ALR$.

Although it degrades the quality of the fits, the inclusion of $\ALR$ generally
does not much affect the values of the parameters for which $\chi^2$ is
minimized.
Neither does it affect the orientation of the error ellipsoid,
and so its inclusion does not change the `optimal' combination of parameters,
$P_i$, which is required to diagonalize the covariance matrix (see point {\it
(4)}
below).
Its strongest influence is on the oblique parameter $S$ which is only driven
further away from zero (see Tables III and VI) by about one standard deviation.
The right-handed $b$ coupling (Tables IV and V) also gets somewhat
shifted in this way.

\topic{(4) `Optimal' Parameters}
Each of Tables III through VIII is divided into two main parts. The first of
these
parts directly gives the central values and one-sigma errors for
the new-physics parameters that can be inferred from the given uncertainties
in the experimental data.  If this information were the entire story, then our
results would indicate that the data in all cases implies that these
nonstandard
couplings are consistent with zero (the SM prediction).

\figure\bellipse
{A fit of the $\zbb$ couplings $\delta g^g_{\sss L,R}$
to the LEP data from the 1995 Winter Conferences. The SM parameters
are chosen for this fit to be $\ss \as = 0.112$, $\ss \mt = 177$ GeV and
$\mH =  300$ GeV. The cross marks the central values
obtainded in the fit and the three solid lines respectively denote the 1--,
2-- and the 4--sigma
error ellipsoids. The SM prediction lies at the origin, $(0,0)$, and so
lies close to the 4--sigma ellipse. If the SLC value for $\ALR$ is
also included in the fit, then the error ellipses are translated so that the
central value lies at the position of the circle.}

This conclusion is misleading, however, because it ignores the correlations
amongst the nonstandard couplings that are imposed by the data. These
correlations can be most simply illustrated for the scenario described in
Table III, for which the only nonstandard quantities are the neutral
current couplings of the $b$ quark, $\delta g^b_{\sss L,R}$. In this
case the correlations which the data imply for these two parameters can
be displayed by plotting the $n$-standard deviation ellipses in the
$\delta g^b_\ssl - \delta g^b_\ssr$ plane. Such a plot, for the data in Table
IVB, is given in Fig.~\bellipse.

\ref\olddata{The LEP Electroweak Working Group, preprint CERN/PPE/94-187.}

This figure displays the results of the fit to the $\zbb$
couplings, $\delta g^b_{\sss L,R}$. In this fit the SM parameters are fixed
to the values $\mH = 300$ GeV, $\mt = 177$ GeV, and $\as = 0.112$, leaving
only the two parameters $\delta \gl^b$ and $\delta \gr^b$ free to float.
The errors quoted in Table IVB for these couplings
correspond to the projection of the one-sigma ellipse in this figure onto each
of the two axes. Even though these projections separately indicate agreement
with zero at the 1.5-$\sigma$ level,  the plot shows that the central values
obtained in the fit are bounded away from the origin (0,0) by more than
4 sigma. (If the left-right asymmetry $A_{LR}$ from SLC
is included this deviation grows to around 5 sigma, although the
quality of the agreement of  the central value of the fit with the data becomes
worse. Using only the previous 1994 LEP data \olddata\ lowers
the deviation to 3.5 sigma.)

When more than two parameters are fit, a similar plot of the
multidimensional error ellipsoid is obviously less useful. As a
result, we display the same information in a different way in
the second parts of Tables III through VII. Here we define
uncorrelated linear combinations, $P_i$, of the new physics parameters.
These combinations diagonalize the covariance
matrix which defines the error ellipsoid. We regard these parameters
as being `optimal' in the sense that they most reliably indicate
the extent to which the new-physics couplings are bounded away
from the origin.  The central values and standard
deviations for these optimal variables are quoted in the second
parts of Tables III -- VII.

\topic{(5) The Preferred Scenario}
The scenario which gives the best fit  to the data (\ie\ which has the lowest
value for $\chi^2_{\rm min}$/d.o.f.) is that of Tables IV,
in which the only new parameters which are entertained are nonstandard
$\zbb$ couplings.  Considering only LEP data, we find $\chi^2_{\rm min}$/d.o.f.
to be in this case 8.4/10, as compared to 12.4/12 for the corresponding
SM fit.  Although this is not overwhelming evidence against the
SM, it is nonetheless suggestive.  Table IVB shows that if $\mt$ is fixed
and a `low' value for $\as $ is assumed to be given by the low-energy data,
then $\chi^2_{\rm min}$/d.o.f.  improves to 8.4/12.

The low $\chi^2_{\rm min}$/d.o.f. value for the fit with nonstandard
$\zbb$ couplings is not simply an instance
of more parameters giving a better fit.  The introduction of different or
additional new couplings also does not
improve the description of the data. The preference for nonstandard
$b$ couplings is driven by the discrepancy between the measured value
for $R_b$ and its SM prediction.  The introduction of nonstandard $\zbb$
couplings is therefore guaranteed to improve the fit, provided that their
contribution to the total hadronic width can somehow be compensated.
Our fits acheive this compensation by having $\as$ take smaller
values than are found in the pure SM fit.\foot\bob{We thank Bob Holdom for
useful conversations on this point.}

\topic{(6) Implications for $\as$}
Our SM fit, as reported in Table II, gives for the QCD coupling the
value $\as = 0.127 \pm 0.004$. This agrees with previous analyses
of the 1994 data \olddata. We find a similar value when the
new physics is assumed to be purely oblique (Table IV). When all
nonstandard $\zff$ couplings are permitted, then $\as$
becomes poorly determined. This is because one combination
of $\as$ and the various new physics couplings tends to drop out of
expressions for the observables. We have therefore fixed $\as$ to be 0.125
in the fit of Table VIII.
It must be noticed that $\chi^2_{\rm min}$/d.o.f.
is also quite high (7.0/3) for this fit, corresponding to a confidence level
of only $7\%$. Such a high value comes about because, although the value of
$\chi^2_{\rm min}$ does not change with the addition of the extra
parameters, the number of degrees of freedom decreases. This
indicates that the additional parameters do not improve the description
of  the data.

\ref\correlations{B. Holdom, \plb{339}{94}{114};
J. Erler and P. Langacker, Preprint UPR-0632T
(hep-ph/9411203) (1994);
G. Altarelli, R. Barbieri and F. Caravaglios, \plb{349}{95}{145}.}

The most interesting case is when the new physics involves nonstandard
couplings to the third generation, such as is reported in Tables IV, V and
VI. In this case $\as$ is only slightly more poorly constrained than in the
SM fit, but has a central value which is significantly smaller. These tables
give $\as = 0.112 \pm 0.009$. This correlation between new
$b$ physics and lower values for $\as$ has also been noticed in the
1994 data \correlations.

\ref\shifman{M. Shifman, \mpla{10}{95}{605}.}

As has recently been emphasized \shifman, the correlation between
nonstandard $\zbb$ couplings and a low value for $\as$ is all the more
interesting
given that such low values are also fairly consistently indicated by other
determinations of $\as$ that are performed away from the $Z$ resonance.
Furthermore, ref.~\shifman\ argues that the QCD scale that is implied
by this lower value for $\as$ ($\Lambda_{{\sss QCD}}\approx 200\ $MeV)
is also required for the success of other methods, such as the operator
product expansion and QCD sum rules.

Since heavy new physics, as is described by an effective-lagrangian such
as ours, does not influence $\as$ as it is determined at low energies, a
reasonable point of view
would be to take the value for $\as$ from the low-energy data
where the contamination from new physics is negligible. Then this value can be
used as a baseline to search for new physics signals in the data at the
$Z$ resonance. Such an approach has the advantage of making it difficult
for new physics to hide in the uncertainties in the determination of $\as$.
Table IVB gives the result of such a fit.  It is tantalizing that one obtains
in this way a deviation from the SM in the $b$-couplings of 4.4 standard
deviations!
\endtopic

\section{Summary}

We conclude that high precision $Z$-pole physics continues to quantitatively
constrain new physics. Furthermore, the theoretical tools now exist to include
a very
general class of new physics in the analysis of the data in a model-independent
way.  The main result from the 1995 Winter Conferences
is that the SM continues to successfully describe $Z$-pole physics, with no
compelling deviations from its
predictions seen in the data.  That is not to say that new physics is thereby
excluded,
however, since the hypothesis that nonstandard $\zbb$ couplings exist provides
a better fit to the data than does the SM. This is due to the continued
discrepancy
between the measured value of $R_b$ and its SM prediction. The new $b$
couplings that are preferred by the data also continue to
bring the inferred values for $\as$ in line with those that are
obtained in lower-energy determinations. Given the hypothesis that nonstandard
neutral-current $b$ couplings exist, and taking $\as$ to be fixed at a low
value,
the latest data bounds the new couplings away from zero at the 4.4 sigma level,
reinforcing the trend also found at a lower level in last year's results.

\bigskip
\centerline{\bf Acknowledgements}
\bigskip

We would like to acknowledge helpful conversations with Bob Holdom and John
Terning. This research was financially supported by the Swiss National
Foundation, NSERC of Canada, FCAR du Qu\'ebec, the Robert A. Welch Foundation,
and the US National Science Foundation, through grant PHY90-09850.

\listrefs

\vfill\eject


$$\vbox{\tabskip=0pt \offinterlineskip
\halign to \hsize{\strut#& #\tabskip 1em plus 2em minus .5em&\hfil#\hfil
&\hfil#\hfil &\hfil#\hfil &#\tabskip=0pt\cr
\noalign{\hrule}\noalign{\smallskip}\noalign{\hrule}\noalign{\medskip}
&& \hfil \hbox{Quantity} & \hfil \hbox{Experimental Value}&
\hfil \hbox{Standard Model Fit} &\cr
\noalign{\medskip}\noalign{\hrule}\noalign{\medskip}
&& $\Mz$ (GeV)  & $91.1887 \pm 0.0022 $  & input & \cr
&& $\Gamma_\ssz$ (GeV)  & $ 2.4971 \pm 0.0033 $  & $2.4979$ &
\cr
&& $\sigma^h_p $ (nb)  & $41.492 \pm 0.081$ & $41.441$ & \cr
&& $R_e=\Gamma_{\rm had}/\Gamma_{e}$ & $20.843 \pm 0.060$
	& $20.783$ & \cr
&& $R_\mu=\Gamma_{\rm had}/\Gamma_{\mu}$  & $20.805 \pm 0.048$
	& $20.783$ & \cr
&& $R_\tau=\Gamma_{\rm had}/\Gamma_{\tau}$  & $20.798 \pm 0.066$
	& $20.783$ & \cr
&& $\AFB{e}$ & $ 0.0154\pm 0.0030$  & $0.0157$ & \cr
&& $\AFB{\mu}$ & $ 0.0160\pm 0.0017$  & $0.0157$ & \cr
&& $\AFB{\tau}$ & $ 0.0209\pm 0.0024$  & $0.0157$ & \cr
&& $A_\tau(P_\tau)$  & $0.140 \pm 0.008$  & $ 0.145$ & \cr
&& $A_e(P_\tau)$  & $ 0.137 \pm 0.009$  & $ 0.145$ & \cr
&& $R_b      $  & $0.2204 \pm 0.0020$ & $ 0.2157 $ & \cr
&& $R_c      $  & $0.1606 \pm 0.0095$ & $ 0.172 $ & \cr
&& $\AFB{b}$  & $0.1015 \pm 0.0036 $  & $0.1015$ & \cr
&& $\AFB{c}$  & $0.0760 \pm  0.0089$  & $ 0.0724$ &\cr
\noalign{\smallskip}
&& $\ALR^0$  & $0.1637 \pm 0.0075$  & $0.145$ &\cr
\noalign{\medskip}\noalign{\hrule}\noalign{\smallskip}\noalign{\hrule}
}}$$
\centerline{{\bf TABLE I}}
\smallskip
\noindent {\eightrm The experimental values for the precision $\ss Z$-pole
observables considered in the present analysis.}
%

\vfill\eject


$$\vbox{\tabskip=0pt \offinterlineskip
\halign to \hsize{\strut#& #\tabskip 1em plus 2em minus .5em&\hfil#\hfil
&\hfil#\hfil &\hfil#\hfil &#\tabskip=0pt\cr
\noalign{\hrule}\noalign{\smallskip}\noalign{\hrule}\noalign{\medskip}
&& \hfil \hbox{Parameter} & \hfil \hbox{LEP Only}&
\hfil \hbox{LEP and SLC} &\cr
\noalign{\medskip}\noalign{\hrule}\noalign{\medskip}
&& $\as$ & $0.127 \pm 0.004$ &  $0.126 \pm 0.004$ &\cr
&& $\mt \hbox{(GeV)}$ & $174 \pm 8$ & $180 \pm 7$ &\cr
\noalign{\medskip}\noalign{\hrule}\noalign{\medskip}
&& $\chi^2_{\rm min}/\hbox{d.o.f.}$ & 12.4/12 (42\% C.L.) &  19.0/13 (12\%
C.L.) &\cr
\noalign{\medskip}\noalign{\hrule}\noalign{\smallskip}\noalign{\hrule}
}}$$
\centerline{{\bf TABLE II: Standard Model}}
\smallskip
\noindent {\eightrm The SM parameters ($\ss \as$ and $\ss \mt$ in GeV)
as determined by fitting to the observables of Table I. }
%


$$\vbox{\tabskip=0pt \offinterlineskip
\halign to \hsize{\strut#& #\tabskip 1em plus 2em minus .5em&\hfil#\hfil
& \hfil#\hfil &\hfil#\hfil & \hfil#\hfil &\hfil#\hfil &#\tabskip=0pt\cr
\noalign{\hrule}\noalign{\smallskip}\noalign{\hrule}\noalign{\medskip}
&& \hbox{Parameter} & \hbox{LEP Only}&\hbox{Pull}&
\hbox{LEP and SLC} &\hbox{Pull  } &\cr
\noalign{\medskip}\noalign{\hrule}\noalign{\medskip}
&& S & -0.08 $\pm$ 0.20 & 0.4 & -0.20 $\pm$ 0.20 &  1.0 &\cr
&& T & -0.10 $\pm$ 0.22 & 0.5 & -0.13 $\pm$ 0.22 &  0.6 &\cr
\noalign{\medskip}\noalign{\hrule}\noalign{\medskip}
&& 0.73 S - 0.68 T & 0.005 $\pm$ 0.085 & 0.1
&-0.068 $\pm$ 0.079 & 0.9 &\cr
&& 0.68 S + 0.73 T & -0.13 $\pm$ 0.29 & 0.5
& -0.23 $\pm$ 0.28 & 0.8 &\cr
\noalign{\medskip}\noalign{\hrule}\noalign{\medskip}
&& $\as$ & 0.127 $\pm$ 0.005 && 0.127 $\pm$ 0.005  &&\cr
\noalign{\medskip}\noalign{\hrule}\noalign{\medskip}
&& $\chi^2_{\rm min}/\hbox{d.o.f.}$ & $12.6/11$ (32\% C.L.) &&  17.6/12
(13\% C.L.)  &
&\cr
\noalign{\medskip}\noalign{\hrule}\noalign{\smallskip}\noalign{\hrule}
}}$$
\centerline{{\bf TABLE III: Oblique Parameters}}
\smallskip
\noindent {\eightrm The oblique parameters $\ss S$ and $\ss T$ and the
SM parameter $\ss \as$ as determined by fitting to the observables of Table I.
$\ss \mt$ is taken to be fixed at 179 GeV. }
%
\vfill\eject


$$\vbox{\tabskip=0pt \offinterlineskip
\halign to \hsize{\strut#& #\tabskip 1em plus 2em minus .5em&\hfil#\hfil
& \hfil#\hfil &\hfil#\hfil & \hfil#\hfil &\hfil#\hfil &#\tabskip=0pt\cr
\noalign{\hrule}\noalign{\smallskip}\noalign{\hrule}\noalign{\medskip}
&& \hbox{Parameter} & \hbox{LEP Only}&\hbox{Pull}&
\hbox{LEP and SLC} &\hbox{Pull  } &\cr
\noalign{\medskip}\noalign{\hrule}\noalign{\medskip}
&& $\delta g^b_\ssl$ & -0.0048 $\pm$ 0.0043 & 1.1&  -0.0037 $\pm$ 0.0043 &
0.9&\cr
&& $\delta g^b_\ssr$ & 0.001 $\pm$ 0.022 & 0.0 & 0.010 $\pm$ 0.022 &  0.5 &\cr
\noalign{\medskip}\noalign{\hrule}\noalign{\medskip}
&& $0.987 \delta g^b_\ssl - 0.160 \delta g^b_\ssr$ & -0.0049 $\pm$ 0.0024 & 2.0
&-0.0052 $\pm$ 0.0024 & 2.2 &\cr
&& $0.160 \delta g^b_\ssl + 0.987 \delta g^b_\ssr$ & 0.000 $\pm$ 0.023 & 0.0
& 0.009 $\pm$ 0.022 & 0.4 &\cr
\noalign{\medskip}\noalign{\hrule}\noalign{\medskip}
&& $\as$ & 0.112 $\pm$ 0.009 && 0.109 $\pm$ 0.009  &&\cr
&& $\mt$ & 177 $\pm$ 9 && 184 $\pm$ 8 &&\cr
\noalign{\medskip}\noalign{\hrule}\noalign{\medskip}
&& $\chi^2_{\rm min}/\hbox{d.o.f.}$ & $8.4/10$ (59\% C.L.) &&  14.1/11 (22\%
C.L.)  &
&\cr
\noalign{\medskip}\noalign{\hrule}\noalign{\smallskip}\noalign{\hrule}
}}$$
\centerline{{\bf TABLE IVA: Nonstandard $\zbb$ Couplings}}
\smallskip
\noindent {\eightrm The new physics parameters $\ss \delta g^b_\ssl$ and $\ss
\delta g^b_\ssr$ and the SM parameters $\ss \as$ and $\ss \mt$ as determined by
fitting to the observables of Table I. }
%


$$\vbox{\tabskip=0pt \offinterlineskip
\halign to \hsize{\strut#& #\tabskip 1em plus 2em minus .5em&\hfil#\hfil
& \hfil#\hfil &\hfil#\hfil & \hfil#\hfil &\hfil#\hfil &#\tabskip=0pt\cr
\noalign{\hrule}\noalign{\smallskip}\noalign{\hrule}\noalign{\medskip}
&& \hbox{Parameter} & \hbox{LEP Only}&\hbox{Pull}&
\hbox{LEP and SLC} &\hbox{Pull  } &\cr
\noalign{\medskip}\noalign{\hrule}\noalign{\medskip}
&& $\delta g^b_\ssl$ & -0.0047 $\pm$ 0.0037 & 1.3 &  -0.0038 $\pm$ 0.0037 &
1.0 &\cr
&& $\delta g^b_\ssr$ & 0.00050 $\pm$ 0.019 & 0.0 & 0.0094 $\pm$ 0.019 &  0.5
&\cr
\noalign{\medskip}\noalign{\hrule}\noalign{\medskip}
&& $0.98 \delta g^b_\ssl - 0.18 \delta g^b_\ssr$ & -0.0048 $\pm$ 0.0011 & 4.4
&-0.0055 $\pm$ 0.0011 & 5.0 &\cr
&& $0.18 \delta g^b_\ssl + 0.98 \delta g^b_\ssr$ & -0.00036 $\pm$ 0.020 & 0.0
& 0.0086 $\pm$ 0.020 & 0.4 &\cr
\noalign{\medskip}\noalign{\hrule}\noalign{\medskip}
&& $\chi^2_{\rm min}/\hbox{d.o.f.}$ & $8.4/12$ (76\% C.L.) &&  14.2/13 (36\%
C.L.)  &
&\cr
\noalign{\medskip}\noalign{\hrule}\noalign{\smallskip}\noalign{\hrule}
}}$$
\centerline{{\bf TABLE IVB: Nonstandard $\zbb$ Couplings}}
\smallskip
\noindent {\eightrm The new physics parameters $\ss \delta g^b_\ssl$ and $\ss
\delta g^b_\ssr$ as determined by fitting to the observables of Table I. In
this fit
we imagine obtaining the SM parameters $\ss \as$ and $\ss \mt$ from experiments
away from the $\ss Z$ pole (lower energies for $\ss \as$ and Fermilab for $\ss
\mt$),
and so fix them to $\ss \as = 0.112$ and $\ss \mt = 177$. }
%

\vfill\eject

$$\vbox{\tabskip=0pt \offinterlineskip
\halign to \hsize{\strut#& #\tabskip 1em plus 2em minus .5em&\hfil#\hfil
& \hfil#\hfil &\hfil#\hfil & \hfil#\hfil &\hfil#\hfil &#\tabskip=0pt\cr
\noalign{\hrule}\noalign{\smallskip}\noalign{\hrule}\noalign{\medskip}
&& \hbox{Parameter} & \hbox{LEP Only}&\hbox{Pull}&
\hbox{LEP and SLC} &\hbox{Pull  } &\cr
\noalign{\medskip}\noalign{\hrule}\noalign{\medskip}
&& $\delta g^b_\ssl$ & -0.0049 $\pm$ 0.0044 & 1.1 &  -0.0036 $\pm$ 0.0043
&  0.8 &\cr
&& $\delta g^b_\ssr$ & 0.000 $\pm$  0.23 & 0.0 & 0.011 $\pm$ 0.022 &  0.5 &\cr
&& $\delta g^\tau_\ssl$ & -0.0002 $\pm$ 0.0011 & 0.2 & 0.0002 $\pm$ 0.0011 &
0.2 &\cr
&& $\delta g^\tau_\ssr$ & -0.0003 $\pm$ 0.0012 & 0.2 & 0.0001 $\pm$ 0.0012 &
0.1 &\cr
\noalign{\medskip}\noalign{\hrule}\noalign{\medskip}
&& P1 & -0.001 $\pm$ 0.023 & 0.0 & 0.010 $\pm$ 0.023 & 0.5 &\cr
&& P2 & -0.0048 $\pm$ 0.0024 & 2.0 & -0.0053 $\pm$ 0.0024 & 2.2 &\cr
&& P3 & -0.0007 $\pm$ 0.0014 & 0.5 & -0.0003 $\pm$ 0.0014 & 0.2 &\cr
&& P4 & -0.00003$\pm$0.00061 & 0.0 & 0.00005 $\pm$ 0.00061 & 0.1 &\cr
\noalign{\medskip}\noalign{\hrule}\noalign{\medskip}
&& $\as$ & 0.112 $\pm$ 0.009 && 0.109 $\pm$ 0.009  &&\cr
&& $\mt$ & 176 $\pm$ 10 && 185 $\pm$ 9 &&\cr
\noalign{\medskip}\noalign{\hrule}\noalign{\medskip}
&& $\chi^2_{\rm min}/\hbox{d.o.f.}$ & $8.3/8$ (40\% C.L.) &&  14.1/9 (12\%
C.L.)  &
&\cr
\noalign{\medskip}\noalign{\hrule}\noalign{\smallskip}\noalign{\hrule}
}}$$
\noindent
$P1 \equiv  0.157 \; \delta g^b_\ssl + 0.987  \; \delta g^b_\ssr +
0.012  \; \delta g^\tau_\ssl + 0.008  \; \delta g^\tau_\ssr$,\hfil\break
$P2 \equiv 0.985  \; \delta g^b_\ssl - 0.156  \; \delta g^b_\ssr -
0.059 \; \delta g^\tau_\ssl -  0.049 \; \delta g^\tau_\ssr$,\hfil\break
$P3 \equiv 0.073 \; \delta g^b_\ssl - 0.026 \; \delta g^b_\ssr +
0.677 \; \delta g^\tau_\ssl + 0.732 \; \delta g^\tau_\ssr$,\hfil\break
$P4 \equiv 0.009 \; \delta g^b_\ssl - 0.004 \; \delta g^b_\ssr +
0.734 \; \delta g^\tau_\ssl - 0.679 \; \delta g^\tau_\ssr$.
\bigskip
\centerline{{\bf TABLE V: Nonstandard Third Generation Couplings}}
\smallskip
\noindent {\eightrm The parameters $\ss \delta g^b_\ssl, \delta g^b_\ssr,
\delta g^\tau_\ssl$ and $\ss \delta g^\tau_\ssr$ and the SM parameters $\ss
\as$ and $\ss \mt$ as determined by fitting to the observables of Table
I.}
%

\vfill\eject

$$\vbox{\tabskip=0pt \offinterlineskip
\halign to \hsize{\strut#& #\tabskip 1em plus 2em minus .5em&\hfil#\hfil
& \hfil#\hfil &\hfil#\hfil & \hfil#\hfil &\hfil#\hfil &#\tabskip=0pt\cr
\noalign{\hrule}\noalign{\smallskip}\noalign{\hrule}\noalign{\medskip}
&& \hbox{Parameter} & \hbox{LEP Only}&\hbox{Pull}&
\hbox{LEP and SLC} &\hbox{Pull  } &\cr
\noalign{\medskip}\noalign{\hrule}\noalign{\medskip}
&& $\delta g^b_\ssl$ & -0.0041 $\pm$ 0.0053 & 0.8 &  -0.0004 $\pm$ 0.0050
&  0.1 &\cr
&& $\delta g^b_\ssr$ & 0.005 $\pm$  0.027 & 0.2 & 0.026 $\pm$ 0.025 & 1.0 &\cr
&& $\delta g^\tau_\ssl$ & -0.0001 $\pm$ 0.0012 & 0.1 & 0.0006 $\pm$ 0.0011 &
0.6 &\cr
&& $\delta g^\tau_\ssr$ & -0.0000 $\pm$ 0.0013 & 0.0 & 0.0008 $\pm$ 0.0013 &
0.6 &\cr
&& S      & -0.10 $\pm$ 0.26 & 0.4 & -0.33 $\pm$ 0.24 & 1.4 &\cr
&& T      & -0.11 $\pm$ 0.23 & 0.5 & -0.16 $\pm$ 0.23 & 0.7 &\cr
\noalign{\medskip}\noalign{\hrule}\noalign{\medskip}
&& P1 & -0.0050 $\pm$ 0.0024 & 2.1 & -0.0050 $\pm$ 0.0024 & 2.1 &\cr
&& P2 & -0.002 $\pm$ 0.020 & 0.1 & 0.002 $\pm$ 0.020 & 0.1 &\cr
&& P3 &  0.00004 $\pm$ 0.00060 & 0.1 & 0.00004 $\pm$ 0.00060 & 0.1 &\cr
&& P4 & -0.0002$\pm$0.0014 & 0.1 & -0.0002 $\pm$ 0.0014 & 0.1 &\cr
&& P5 & -0.14 $\pm$ 0.33 & 0.4 & -0.35 $\pm$ 0.31 & 1.1  &\cr
&& P6 & -0.02 $\pm$ 0.12 & 0.2 & 0.12 $\pm$ 0.11 & 1.1 &\cr
\noalign{\medskip}\noalign{\hrule}\noalign{\medskip}
&& $\as$ & 0.112 $\pm$ 0.009 && 0.110 $\pm$ 0.009  &&\cr
\noalign{\medskip}\noalign{\hrule}\noalign{\medskip}
&& $\chi^2_{\rm min}/\hbox{d.o.f.}$ & $8.2/7$ (32\% C.L.) &&  12.2/8
(14\% C.L.)  &&\cr
\noalign{\medskip}\noalign{\hrule}\noalign{\smallskip}\noalign{\hrule}
}}$$
\noindent
$P1 \equiv  0.985 \; \delta g^b_\ssl - 0.172 \; \delta g^b_\ssr + 0.00045  \; S
- 0.00034  \; T$, \hfil\break
$P2 \equiv 0.171 \; \delta g^b_\ssl + 0.976 \; \delta g^b_\ssr + 0.109  \; S
 - 0.081  \; T $, \hfil\break
$P3 \equiv  0.00014  \; S
- 0.00094  \; T  + 0.757 \; \delta g^\tau_\ssl - 0.653 \; \delta g^\tau_\ssr$,
\hfil\break
$P4 \equiv  0.00560  \; S
 - 0.00399  \; T   + 0.653 \; \delta g^\tau_\ssl + 0.757 \; \delta
g^\tau_\ssr$,
\hfil\break
$P5 \equiv  -0.01 \; \delta g^b_\ssl - 0.03 \; \delta g^b_\ssr + 0.768  \; S
 + 0.639  \; T$, \hfil\break
$P6 \equiv  0.02 \; \delta g^b_\ssl + 0.13 \; \delta g^b_\ssr - 0.631  \; S
 + 0.765  \; T$.
\bigskip
\centerline{{\bf TABLE VI: Nonstandard Third Generation and Oblique Couplings}}
\smallskip
\noindent {\eightrm The parameters $\ss \delta g^b_\ssl, \delta
g^b_\ssr, S$ and $\ss T$ and the SM parameter $\ss \as$
as determined by fitting to the observables of Table I. $\ss \mt$ is
taken to be fixed at $179 \GeV$.}
%

\vfill\eject

$$\vbox{\tabskip=0pt \offinterlineskip
\halign to \hsize{\strut#& #\tabskip 1em plus 2em minus .5em&\hfil#\hfil
& \hfil#\hfil &\hfil#\hfil & \hfil#\hfil &\hfil#\hfil &#\tabskip=0pt\cr
\noalign{\hrule}\noalign{\smallskip}\noalign{\hrule}\noalign{\medskip}
&& \hbox{Parameter} & \hbox{LEP Only}&\hbox{Pull}&
\hbox{LEP and SLC} &\hbox{Pull  } &\cr
\noalign{\medskip}\noalign{\hrule}\noalign{\medskip}
&& $\delta g^e_\ssl$ & -0.000873 $\pm$ 0.000938 & 0.9 & -0.00144 $\pm$ 0.000899
& 1.6 &\cr
&& $\delta g^e_\ssr$ & -0.000574 $\pm$ 0.000768 & 0.7 & -0.00119 $\pm$ 0.000712
&  1.7 &\cr
&& $\delta g^\mu_\ssl$ & -0.000787 $\pm$ 0.00207 & 0.4 & -0.00035 $\pm$ 0.00206
& 0.2 &\cr
&& $\delta g^\mu_\ssr$ & -0.000518 $\pm$ 0.00226 & 0.2 & 0.0000348 $\pm$
0.00224
&0.0 &\cr
\noalign{\medskip}\noalign{\hrule}\noalign{\medskip}
&& P1 & 0.00105 $\pm$ 0.00119 & 0.9 & 0.00186 $\pm$ 0.00113 & 1.7 &\cr
&& P2 & -0.000928 $\pm$ 0.00301 & 0.3 & -0.000266 $\pm$ 0.00299 & 0.1 &\cr
&& P3 & 0.0000701 $\pm$ 0.000317 & 0.2 & -0.0000102 $\pm$ 0.000315 & 0.0 &\cr
&& P4 & -0.0000132 $\pm$ 0.000517 & 0.0 & -0.00028 $\pm$ 0.000507 & 0.6 &\cr
\noalign{\medskip}\noalign{\hrule}\noalign{\medskip}
&& $\as$ & 0.130 $\pm$ 0.006 && 0.131 $\pm$ 0.006  &&\cr
&& $\mt$ & 163 $\pm$ 15 && 162 $\pm$ 15 &&\cr
\noalign{\medskip}\noalign{\hrule}\noalign{\medskip}
&& $\chi^2_{\rm min}/\hbox{d.o.f.}$ & $11.4/8$ (18\% C.L.) &&  15.9/9 (7\%
C.L.)  &
&\cr
\noalign{\medskip}\noalign{\hrule}\noalign{\smallskip}\noalign{\hrule}
}}$$
\noindent
$P1 \equiv  -0.766 \delta g^e_\ssl - 0.589 \delta g^e_\ssr - 0.18 \delta
g^\mu_\ssl +
0.183 \delta g^\mu_\ssr$  ,\hfil\break
$P2 \equiv 0.0221 \delta g^e_\ssl - 0.00545 \delta g^e_\ssr + 0.673 \delta
g^\mu_\ssl +
0.74 \delta g^\mu_\ssr$,\hfil\break
$P3 \equiv -0.642 \delta g^e_\ssl + 0.683 \delta g^e_\ssr + 0.27 \delta
g^\mu_\ssl -
0.221 \delta g^\mu_\ssr  $,\hfil\break
$P4 \equiv -0.0301 \delta g^e_\ssl + 0.431 \delta g^e_\ssr - 0.665 \delta
g^\mu_\ssl +
0.609 \delta g^\mu_\ssr$.
\bigskip
\centerline{{\bf TABLE VII: Nonstandard Lepton Couplings}}
\smallskip
\noindent {\eightrm The parameters $\ss \delta g^e_\ssl, \delta g^e_\ssr,
\delta g^\mu_\ssl$ and $\ss \delta g^\mu_\ssr$ and the SM parameters $\ss
\as$ and $\ss \mt$ as determined by fitting to the observables of Table
I. }
%

\vfill\eject
$$\vbox{\tabskip=0pt \offinterlineskip
\halign to \hsize{\strut#& #\tabskip 1em plus 2em minus .5em&\hfil#\hfil
& \hfil#\hfil &\hfil#\hfil & \hfil#\hfil &\hfil#\hfil &#\tabskip=0pt\cr
\noalign{\hrule}\noalign{\smallskip}\noalign{\hrule}\noalign{\medskip}
&& \hbox{Parameter} & \hbox{LEP Only}&\hbox{Pull}&
\hbox{LEP and SLC} &\hbox{Pull  } &\cr
\noalign{\medskip}\noalign{\hrule}\noalign{\medskip}
&& $\delta g^e_\ssl$ & 0.00005 $\pm$ 0.00088 & 1.0 &  -0.00102 $\pm$ 0.00066
&  1.5 &\cr
&& $\delta g^e_\ssr$ & 0.00002 $\pm$ 0.00099 & 0.0 & -0.00120 $\pm$ 0.00073 &
1.6 &\cr
&& $\delta g^\mu_\ssl$ & 0.0000 $\pm$ 0.0021 & 0.0 & 0.0010 $\pm$ 0.0020 &
0.5 &\cr
&& $\delta g^\mu_\ssr$ & -0.0002 $\pm$ 0.0024 & 0.1 & 0.0010 $\pm$ 0.0023 &
0.4 &\cr
&& $\delta g^\tau_\ssl$ & -0.0001 $\pm$ 0.0010 & 0.1 & 0.0001 $\pm$ 0.0010 &
0.1 &\cr
&& $\delta g^\tau_\ssr$ & -0.0002 $\pm$ 0.0011 & 0.2 & -0.0001 $\pm$ 0.0011 &
0.1 &\cr
&& $\delta g^b_\ssl$ & -0.0048 $\pm$ 0.0066 & 0.7 & 0.0017 $\pm$ 0.0056 &
0.3 &\cr
&& $\delta g^b_\ssr$ & 0.000 $\pm$ 0.034 & 0.0 & 0.035 $\pm$ 0.028 & 1.2 &\cr
&& $\delta g^c_\ssl$ &-0.008 $\pm$ 0.013 & 0.6 & -0.012 $\pm$ 0.012 & 1.0 &\cr
&& $\delta g^c_\ssr$ & 0.013 $\pm$ 0.022 & 0.6 &  0.003 $\pm$ 0.021 & 0.2 &\cr
&& $\delta_{{\sss UD}}$ &
                    0.0029 $\pm$ 0.0038 & 0.8 & 0.0029 $\pm$ 0.0038 & 0.8 &\cr
\noalign{\medskip}\noalign{\hrule}\noalign{\medskip}
&& P1 & -0.004 $\pm$ 0.037 & 0.1 & 0.032 $\pm$ 0.031 & 1.0 &\cr
&& P2 & -0.009 $\pm$ 0.020 & 0.5 & -0.015 $\pm$ 0.019 & 0.8 &\cr
&& P3 &  0.012 $\pm$ 0.011 & 1.1 & 0.011 $\pm$ 0.011 & 1.0 &\cr
&& P4 & 0.0001 $\pm$ 0.0029  & 0.0 & 0.0007 $\pm$ 0.0029 & 0.3 &\cr
&& P5 & -0.0014 $\pm$ 0.0020 & 2.1 & -0.0041 $\pm$ 0.0020 & 2.1  &\cr
&& P6 &  0.0002 $\pm$ 0.0014 & 0.1 & -0.0001 $\pm$ 0.0014 & 0.1 &\cr
&& P7 & -0.00018$\pm$ 0.00080 & 0.2& -0.00019 $\pm$ 0.00080 &  0.2 &\cr
&& P8 & -0.00002$\pm$ 0.00072 & 0.0& -0.00076 $\pm$ 0.00063 & 1.2 &\cr
&& P9 & -0.00005 $\pm$ 0.00027 & 0.2 & -0.00005 $\pm$ 0.00027 & 0.2 &\cr
&& P10& 0.00011$\pm$ 0.00038 & 0.3  & 0.00011  $\pm$ 0.00038 & 0.3 &\cr
&& P11& -0.00001$\pm$ 0.00052 & 0.0 &-0.00001 $\pm$ 0.00052 & 0.0 &\cr
\noalign{\medskip}\noalign{\hrule}\noalign{\medskip}
&& $\chi^2_{\rm min}/\hbox{d.o.f.}$ & $7.0/3$ (7\% C.L.) &&  10.4/4
(3\% C.L.)  &&\cr
\noalign{\medskip}\noalign{\hrule}\noalign{\smallskip}\noalign{\hrule}
}}$$
\noindent
$P1 \equiv  0.17 \; \delta g^b_\ssl + 0.91 \; \delta g^b_\ssr - 0.15 \; \delta
g^c_\ssl
 - 0.34 \; \delta g^c_\ssr - 0.02 \; \delta g^e_\ssl - 0.02 \; \delta g^e_\ssr
 + 0.02 \; \delta g^\mu_\ssl + 0.02 \; \delta g^\mu_\ssr$, \hfil\break
$P2 \equiv  -0.05 \; \delta g^b_\ssl - 0.37 \; \delta g^b_\ssr - 0.26 \; \delta
g^c_\ssl
 - 0.89 \; \delta g^c_\ssr + 0.01 \; \delta g^e_\ssl + 0.01 \; \delta g^e_\ssr
 - 0.01 \; \delta g^\mu_\ssl - 0.01 \; \delta g^\mu_\ssr - 0.04 \; \dUD$,
\hfil\break
$P3 \equiv -0.06 \; \delta g^b_\ssl - 0.04 \; \delta g^b_\ssr - 0.90 \; \delta
g^c_\ssl
 + 0.27 \; \delta g^c_\ssr + 0.33 \; \dUD$, \hfil\break
$P4 \equiv 0.01 \; \delta g^b_\ssl + 0.03 \; \delta g^b_\ssr + 0.05 \; \delta
g^e_\ssl
+ 0.05 \; \delta g^e_\ssr
 - 0.65 \; \delta g^\mu_\ssl - 0.75 \; \delta g^\mu_\ssr - 0.01 \; \delta
g^\tau_\ssl
 - 0.02 \; \delta g^\tau_\ssr $, \hfil\break
$P5 \equiv 0.92 \; \delta g^b_\ssl - 0.17 \; \delta g^b_\ssr + 0.06 \; \delta
g^c_\ssl
 - 0.02 \; \delta g^c_\ssr + 0.01 \; \delta g^e_\ssl + 0.01 \; \delta
g^\mu_\ssl + 0.01
\; \delta g^\tau_\ssl
 - 0.01 \; \delta g^\tau_\ssr + 0.35 \; \dUD$, \hfil\break
$P6 \equiv 0.03 \; \delta g^e_\ssl + 0.03 \; \delta g^e_\ssr
 + 0.02 \; \delta g^\mu_\ssl + 0.02 \; \delta g^\mu_\ssr - 0.65 \; \delta
g^\tau_\ssl
 - 0.76 \; \delta g^\tau_\ssr $, \hfil\break
$P7 \equiv -0.22 \; \delta g^b_\ssl + 0.04 \; \delta g^b_\ssr + 0.20 \; \delta
g^c_\ssl
 - 0.09 \; \delta g^c_\ssr - 0.15 \; \delta g^e_\ssl + 0.16 \; \delta g^e_\ssr
 - 0.31 \; \delta g^\mu_\ssl + 0.27 \; \delta g^\mu_\ssr - 0.44 \; \delta
g^\tau_\ssl
 + 0.38 \; \delta g^\tau_\ssr + 0.59 \; \dUD$, \hfil\break
$P8 \equiv 0.03 \; \delta g^b_\ssr + 0.66 \; \delta g^e_\ssl + 0.75 \; \delta
g^e_\ssr
 + 0.05 \; \delta g^\mu_\ssl + 0.05 \; \delta g^\mu_\ssr + 0.04 \; \delta
g^\tau_\ssl
 + 0.02 \; \delta g^\tau_\ssr - 0.02 \; \dUD$, \hfil\break
$P9 \equiv 0.14 \; \delta g^b_\ssl - 0.02 \; \delta g^b_\ssr - 0.11 \; \delta
g^c_\ssl
 + 0.05 \; \delta g^c_\ssr - 0.70 \; \delta g^e_\ssl + 0.61 \; \delta g^e_\ssr
 - 0.08 \; \delta g^\mu_\ssl + 0.07 \; \delta g^\mu_\ssr - 0.01 \; \delta
g^\tau_\ssl
 + 0.01 \; \delta g^\tau_\ssr - 0.32 \; \dUD$, \hfil\break
$P10 \equiv -0.17 \; \delta g^b_\ssl + 0.03 \; \delta g^b_\ssr + 0.15 \; \delta
g^c_\ssl
 - 0.06 \; \delta g^c_\ssr - 0.22 \; \delta g^e_\ssl + 0.19 \; \delta g^e_\ssr
 + 0.62 \; \delta g^\mu_\ssl - 0.54 \; \delta g^\mu_\ssr + 0.06 \; \delta
g^\tau_\ssl
 - 0.06 \; \delta g^\tau_\ssr + 0.42 \; \dUD$, \hfil\break
$P11 \equiv 0.15 \; \delta g^b_\ssl - 0.03 \; \delta g^b_\ssr - 0.13 \; \delta
g^c_\ssl
 + 0.06 \; \delta g^c_\ssr + 0.11 \; \delta g^e_\ssl - 0.09 \; \delta g^e_\ssr
 + 0.29 \; \delta g^\mu_\ssl - 0.25 \; \delta g^\mu_\ssr - 0.61 \; \delta
g^\tau_\ssl
 + 0.53 \; \delta g^\tau_\ssr - 0.37 \; \dUD$.
\bigskip
\centerline{{\bf TABLE VIII: Nonstandard $\zff$ Couplings}}
\medskip
\noindent {\eightrm The parameters $\ss \delta g^f_\ssl$ and $\ss \delta
g^f_\ssr$ as determined by fitting to the observables of Table I.
$\ss \mt$ and $\ss \as$ are taken to be fixed at 179 GeV and 0.125
respectively. The
combination $\ss \dUD$ is the combination of light-quark neutral-current
couplings
defined in the text. }
%

\figurecaptions

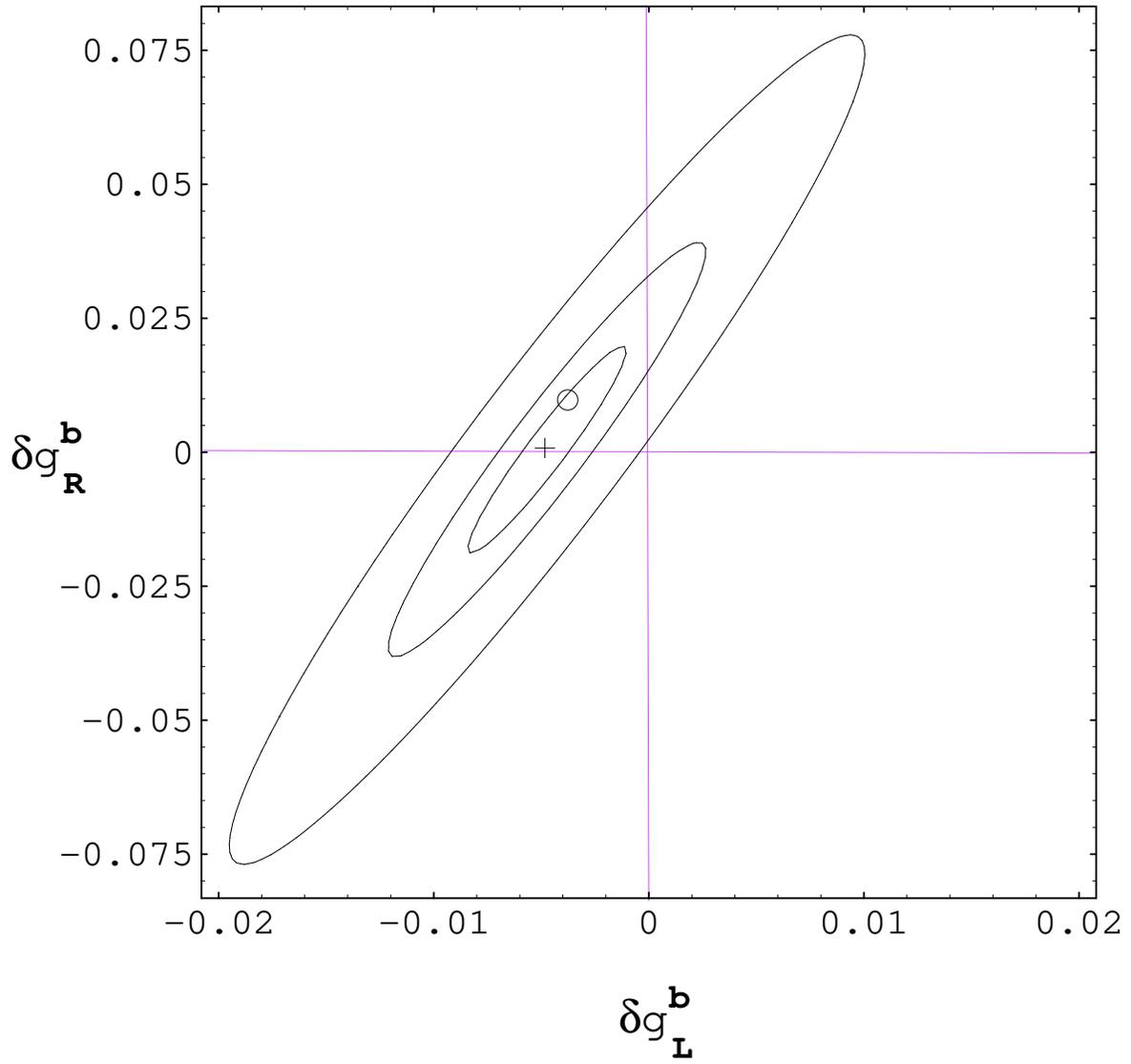
\bye